Adaptive sampling dual terahertz comb spectroscopy using free-running dual femtosecond lasers

Takeshi Yasui,[1,2,3] Ryuji Ichikawa,[1] Yi-Da Hsieh,[1,3] Kenta Hayashi,[1] Francis Hindle,[4] Yoshiyuki Sakaguchi,[2] Kaoru Minoshima,[3,5] and Hajime Inaba[3,6]

[1]Institute of Technology and Science, Tokushima University, 2-1 Minami-Josanjima, Tokushima 770-8506, Japan

[2]Graduate School of Engineering Science, Osaka University, 1-3 Machikaneyama, Toyonaka, Osaka 560-8531, Japan

[3]JST, ERATO, MINOSHIMA Intelligent Optical Synthesizer Project, 2-1 Minami-Josanjima, Tokushima 770-8506, Japan

[4]Laboratoire de Physico-Chimie de l'Atmosphère, Université du Littoral Côte d'Opale, 189A Av. Maurice Schumann, Dunkerque 59140, France

[5]Graduate School of Informatics and Engineering, The University of Electro-Communications, 1-5-1, Chofugaoka, Chofu, Tokyo 182-8585, Japan

[6]National Metrology Institute of Japan, National Institute of Advanced Industrial

Science and Technology, 1-1-1 Umezono, Tsukuba, Ibaraki 305-8563, Japan

Abstract

Dual terahertz (THz) comb spectroscopy is a promising methods for high accuracy, high resolution, and broadband THz spectroscopy because the mode-resolved THz comb spectrum possesses both characteristics of broadband THz radiation and narrow-linewidth continuous-wave THz radiation and all frequency mode of THz comb can be phase-locked to a microwave frequency standard. However, requirement of stabilized dual femtosecond lasers has often hindered wide use of this method. In this article, we demonstrated the adaptive sampling, dual THz comb spectroscopy, enabling use of free-running dual femtosecond lasers. To correct the non-linearity of time and frequency scale caused by the laser timing jitter, an adaptive sampling clock is generated by dual THz-comb-referenced spectrum analysers and is used for a timing signal in a data acquisition board. The demonstrated results did not only indicate the implementation of dual THz comb spectroscopy with free-running dual lasers but also implied the superiority of its spectroscopic performance over the dual THz comb spectroscopy with stabilized dual lasers.

# 1. Introduction

Terahertz (THz) radiation, lying between 0.1 and 10 THz, has attracted many attentions as a new tool for material characterization and sensing because one can observed interested phenomena characteristic in this region, such as optical phonon scattering and plasma frequency of solid, dielectric property including ionic polarization and orientation polarization, rotational transition of polar gas molecules, and intermolecular interaction [1]. Therefore, the spectroscopic analysis has been the main driving force in THz technology and science. To probe a variety of phenomena in THz region and characterise them precisely, a spectroscopic technique with high accuracy, high resolution, and wide spectral coverage is strongly desired. Unfortunately, all of these requirements can not be satisfied at the same time using standard THz spectroscopic techniques including THz time-domain spectroscopy (THz-TDS) with coherent broadband THz radiation [2], far-infrared Fourier transform spectroscopy with incoherent broadband THz radiation [3], and THz frequency-domain spectroscopy (THz-FDS) with tunable continuous-wave THz (CW-THz) radiation [4].

Recently, THz frequency comb has unlocked further potential for high accuracy, high resolution, and broadband THz spectroscopy [5-7]. THz spectroscopy based on THz comb can combine advantage of THz-TDS and THz-FDS, namely broadband coverage and high resolution, because the mode-resolved THz comb spectrum includes both characteristics of broadband THz radiation and narrow-line CW-THz radiation. Furthermore, the absolute accuracy of all frequency modes in THz comb is secured by phase-locking of THz comb to a microwave frequency standard via laser control. The mode-resolved THz comb spectrum can be obtained by dual comb techniques in time [7-11] and frequency domains [5, 6]. Since the resulting spectrum is composed of a series of frequency spike with a frequency spacing equal to a mode-locked frequency and a linewidth equal to an inverse of an observation window of temporal waveform, each comb mode can be used as a universal frequency scale for broadband THz radiation. Furthermore, the combination of the spectral interleaving technique [12, 13] with dual comb spectroscopy has made the comb-mode-resolved spectrum gapless, leading to large enhancement of spectral resolution and accuracy [14]. In these previous researches for dual THz comb spectroscopy, precise control of dual femtosecond lasers is essential. However, use of such the specially stabilized dual lasers has often hindered the dual THz comb spectroscopy from spreading out in various applications even though its spectroscopic performance is sufficiently high. If such the dual THz comb spectroscopy can be implemented using *free-running*, or *unstabilized*, dual lasers, the application scope of the dual THz comb spectroscopy will be largely extended.

More recently, the adaptive sampling technique has been proposed for the dual comb spectroscopy in visible [15] and infrared regions [16], enabling to use free-running dual lasers by recovering the nonlinearity of the time axis using the adaptive sampling clock. However, there have been no attempts to apply the adaptive sampling technique for dual THz comb spectroscopy. In this article, we report dual THz comb spectroscopy with free-running dual lasers by modifying the adaptive sampling technique for THz comb. Dual THz-comb-referenced spectrum analysers [17-22] are effectively used to generate an adaptive sapling clock for dual THz comb spectroscopy.

2. Principle

Figure 1(a) shows a signal flowchart of dual THz comb spectroscopy. In the dual THz comb spectroscopy, a temporal waveform of THz pulse train is acquired over many repetition periods using an asynchronous optical sampling (ASOPS) method with two mode-locked femtosecond lasers (repetition freq. = $f_{rep1}$ and $f_{rep2}$) with a frequency offset between them (= $f_{offset}$ = $f_{rep2}$ - $f_{rep1}$) [22-25]. As a result of ASOPS, a repetition period of THz pulse train (= $1/f_{rep1}$) is expanded to $1/f_{offset}$ based on a temporal magnification factor (TMF) of $f_{rep1}/f_{offset}$. The temporal signal slow-downed to RF region, namely RF pulse train, can be directly measured in a data acquisition board of computer without the need for mechanical time-delay scanning. Then, the comb-mode-resolved spectrum is obtained in RF region, namely RF comb, by calculating Fourier transform (FT) of RF pulse train. Finally, THz comb can be obtained by calibrating the frequency scale of RF comb with TMF. If we use free-running dual lasers in ASOPS, a timing jitter between them fluctuates TMF every moment and hence distorts the linearity of time axis in RF pulse train. Such nonlinearity of the time scale propagates the frequency scale of RF comb and THz comb via Fourier transform and the frequency calibration, leading to a large drop of spectral resolution and accuracy.

When the temporal waveform of the distorted RF pulse train is acquired in synchronization with a constant sampling clock using the data acquisition board (namely, constant sampling method), the time scale of the acquired signal is remained distorted as shown in the left part of Fig. 1(b). However, if the temporal waveform of the distorted RF pulse train is acquired in synchronization with an adaptive sampling clock (namely, adaptive sampling method), reflecting fluctuation of TMF by the timing jitter, the linearity of the time scale in RF pulse train can be recovered as shown the right part of Fig. 1(b) [15, 16].

We here consider which of $f_{rep1}$ and $f_{offset}$ is more dominant in uncertainty of TMF, namely timing jitter. The propagation of uncertainty for $f_{rep1}$ and $f_{offset}$ to TMF is given as follows

$$\sigma_{TMF} = \sqrt{\left(\frac{\partial TMF}{\partial f_{rep1}}\right)^2 \left(\sigma_{f_{rep1}}\right)^2 + \left(\frac{\partial TMF}{\partial f_{offset}}\right)^2 \left(\sigma_{f_{offset}}\right)^2}$$
$$= \sqrt{\left(\frac{1}{f_{offset}}\right)^2 \left(\sigma_{f_{rep1}}\right)^2 + \left[\frac{-f_{rep1}}{(f_{offset})^2}\right]^2 \left(\sigma_{f_{offset}}\right)^2} \quad (1)$$
$$= \sqrt{\left(\frac{1}{50}\right)^2 \left(\sigma_{f_{rep1}}\right)^2 + \left[\frac{-100,000,000}{50^2}\right]^2 \left(\sigma_{f_{offset}}\right)^2}$$
$$= \sqrt{4 \times 10^{-4} \times \left(\sigma_{f_{rep1}}\right)^2 + 1.6 \times 10^9 \times \left(\sigma_{f_{offset}}\right)^2}$$

where $\sigma_{TMF}$, $\sigma_{f_{rep1}}$, and $\sigma_{f_{offset}}$ are the standard uncertainty for TMF, $f_{rep1}$, and $f_{offset}$, respectively. Since $\sigma_{f_{offset}}$ is approximate to the root-sum-square value of $\sigma_{f_{rep1}}$ and $\sigma_{f_{rep2}}$ and hence is comparable to $\sigma_{f_{rep1}}$, uncertainty of TMF is mainly due to that of $f_{offset}$ rather than $f_{rep1}$ from Eq. (1). Therefore, the adaptive sampling clock for dual THz comb spectroscopy should be generated by considering the fluctuation of $f_{offset}$ in the frequency range of THz comb spectrum, namely the higher-order harmonic component of $f_{offset}$. This component can be extracted from the higher-order harmonic components of $f_{rep1}$ and $f_{rep2}$ lying around a few THz. To this end, we extracted a beat signal between dual THz comb modes by expanding the THz-comb-referenced spectrum analyser [17, 18] into dual configuration. Figures 2(a) and (b) shows the spectral behavior of this method and the corresponding experimental setup. When two femtosecond laser lights after the wavelength conversion ($f_{rep1}$ = 100 MHz, $f_{rep2}$ = 100,000,050 Hz, and $f_{offset}$ = 50 Hz) are respectively incident onto a pair of photoconductive antenna (PCA1 and PCA2), the photocarrier THz combs with a frequency spacing of $f_{rep1}$ and $f_{rep2}$ (PC-THz comb 1 and 2) are respectively induced in PCA1 and PCA2, respectively. Then, a CW-THz wave from an active frequency multiplier chain (output freq. = $f_{THz}$ = 0.1 THz, linewidth < 0.6 Hz, average power = 2.5 mW) combined with a microwave frequency synthesizer is incident onto both PCA1 and PCA2. As a result of the photoconductive mixing in PCA1 and PCA2, a pair of beat signals between CW-THz wave and PC-THz comb 1 and 2 is generated in RF region. When $m$ (= 1,000) is the order of PC-THz comb mode nearest in frequency to CW-THz wave, frequencies of those two beat signals ($f_{beat1}$ and $f_{beat2}$) are given as $f_{THz} - mf_{rep1}$ and $f_{THz} - mf_{rep2}$, respectively. By the first frequency multiplying (multiplication factor = $N_1$ = 40), electrically mixing between them, low-pass filtering, and the second frequency multiplying (multiplication factor = $N_2$ = 40), we extracted a beat signal between modes of THz comb in THz region as the higher-order component of $f_{offset}$ (freq. = $mN_1N_2f_{offset}$ = 2 MHz). Figure 2(c) shows the spectrum of the extracted signal, which corresponds to a beat signal between modes of THz comb at 4 THz. This signal has sufficiently high frequency and narrow linewidth to use for the adaptive sampling clock in dual THz comb spectroscopy. It is important to note that only one kind of adaptive clock is required for dual THz comb spectroscopy because THz comb is a harmonic frequency comb of $f_{rep1}$ without carrier-envelope-offset frequency ($f_{ceo}$). Such the simplified generation of the adaptive sampling clock will be one advantage of this method over the adaptive sampling dual optical comb spectroscopy, where two kinds of adaptive sampling clocks are required to correct fluctuation of both $f_{rep1}$ and $f_{ceo}$ [15, 16].

3. Results

Figure 3 illustrates a schematic diagram of the experimental setup, which contains dual mode-locked Er-doped fibre lasers ($f_{rep1}$ = 100,000,000 Hz, $f_{rep2}$ = 100,000,050 Hz, and $f_{offset}$ = 50 Hz), a THz optical setup, and a low-pressure gas cell (length = 500 mm, diameter = 40 mm). The temporal waveform of RF pulse train was acquired within a time window size of 200 ms with a digitizer (sampling rate = $2 \times 10^6$ samples/s, the number of sampling point = 400,000, resolution = 20 bit) by using the adaptive sampling clock as a timing signal of the data acquisition in the digitizer. This data acquisition condition at TMF of 2,000,000 (= $f_{rep1}/f_{offset}$) is equivalent to the condition that the temporal waveform of 10 consecutive THz pulses (repetition period = 10 ns) was acquired at a sampling interval of 100 fs with a time window of 100 ns. We acquired temporal waveforms for 10 consecutive THz pulses at a scan rate of 5 Hz and accumulated them to obtain good signal-to-noise ratio. Finally, the amplitude and power spectrum of the THz comb was obtained by Fourier transform and the frequency calibration.

To investigate effect of the proposed method in time domain, we integrated 10,000 temporal waveforms for 10 consecutive THz pulses and compared the integrated temporal waveform between the constant sampling method and the adaptive sampling method. Figures 4(a), (b), and (c) shows a comparison of the integrated temporal waveform among the constant sampling method with free-running dual lasers, the constant sampling method with stabilized dual lasers, and the adaptive sampling method with free-running dual lasers. In Fig. 4(a), the signal of THz pulse train almost disappeared except the first THz pulse. This is because the fluctuated TMF changes temporal position of 10 consecutive THz pulses at every data acquisition, and hence the integration of them makes the signal be buried in the noise. In this way, the constant sampling method can not be used with free-running dual lasers. In Fig. 4(b), 10 consecutive THz pulses were observed even after the signal integration, which is similar to the previous researches of the constant sampling method with stabilized dual lasers [7]. However, peak amplitude of the pulsed THz electric field varied with each pulse. This result implies that the remaining timing jitter fluctuates the relative timing between THz pulse and the probe pulse even though $f_{rep1}$ and $f_{rep2}$ are stabilized by the control system. In Fig. 4(c), nevertheless the dual lasers are in free-running operation, the 10 consecutive THz pulse clearly appears in the integrated temporal waveform. Furthermore, it should be emphasized that peak amplitude of THz electric field was constant for all pulses in contrast to Fig.

4(b). That is to say, the adaptive sampling method seems to be more powerful than the constant sampling method with stabilized dual lasers from the viewpoint of the reduction of timing jitter.

To investigate the effect of the proposed method in frequency domain, we obtained THz comb spectrum by calculating FT of the temporal waveform in Figs. 4(b) and (c). Figures 4(d) and (e) show a comparison of the mode-resolved THz comb spectra between the constant sampling method with stabilized dual lasers and the adaptive sampling method with free-running dual lasers. The spectral envelops of these two THz combs are identical to each other, revealing the periodic modulation due to the multiple reflections in PCA and the sharp spectral dips caused by the atmospheric water vapor. To observe the detailed mode structure into the THz comb, we expanded the spectral region around 0.5 THz. As shown in insets of Figs. 4(d) and (e), the THz comb modes had a frequency spacing of 100 MHz. However, the linewidth of the comb mode in Fig. 4(d) (= 17 MHz) were larger than that in Fig. 4(e) (= 10 MHz). Furthermore, the contrast of the amplitude between the comb mode and comb gap are different between them. In other words, the comb modes in the inset of Figs. 4(d) became blurred whereas those in the inset of Figs. 4(e) maintain the distinct spike. Such the increased linewidth and decreased contrast will leads to decrease of spectral resolution as well as dynamic range and signal-to-noise ratio of amplitude. In this way, the adapting sampling method does not only enables use of free-running lasers but also bring the potential to enhance the spectroscopic performance over the conventional dual THz comb spectroscopy based on the constant sampling method with stabilized dual lasers.

THz spectroscopic analysis of low-pressure molecular gas is one of interesting applications that require the highest resolution and accuracy in the broadband THz spectral range. We here evaluated the potential of the proposed system for high accuracy, high resolution, and broadband THz spectroscopy by measuring low-pressure molecular gasses. To evaluate the spectroscopic performance for an isolated, sharp absorption line, the absorption spectrum of the rotational transition $1_{10} \leftarrow 1_{01}$ of water vapor (freq. = 0.556936 THz) was measured at a gas pressure of 2570 Pa. We estimated the pressure broadening linewidth of this absorption line to be 200 MHz by the pressure-broadening coefficient [26]. Figures 5(a) and (b) shows a comparison of the absorbance spectrum between the constant sampling method with stabilized dual lasers and the adaptive sampling method with free-running dual lasers. By fitting a Lorentzian function to the measured spectral profile, the center frequency and the spectral linewidth of the absorption line were determined to be 0.55698 THz and 196 MHz for Fig. 5(a) and 0.55692 THz and 283 MHz for Fig. 5(b), respectively. The deviation of the determined center frequency from the spectral database values [27] were 44 MHz for Fig. 5(a) and 16 MHz for Fig. 5(b) whereas the determined linewidth were reasonably agreement with the expected pressure-broadening linewidth (= 200 MHz). This comparison indicated that the adaptive sampling dual THz comb spectroscopy with free-running lasers could achieve the spectroscopic performance similar to the dual THz comb spectroscopy with stabilized lasers in the low-pressure gas spectroscopy.

Finally, to demonstrate the capacity to simultaneously probe multiple absorption lines of low-pressure molecular gas, we performed gas-phase spectroscopy of acetonitrile ($CH_3CN$). Since $CH_3CN$ is not only a very abundant species in the interstellar medium but also one of volatile organic compounds in the atmosphere, its spectroscopic analysis is important in the astronomy and the atmospheric pollution. As $CH_3CN$ has a rotational constant $B$ of 9.2 GHz in molecular structure of symmetric top, many groups of rotational transitions regularly spaced by $2B$ (= 18.4 GHz) appears in THz region as shown in Fig. 6(a) [28]. Figures 6(b) and (c) shows a comparison of the absorbance spectrum of $CH_3CN$ at 1 kPa between the constant sampling method with stabilized lasers and the adaptive sampling method with free-running lasers. Many groups of the rotational transitions regularly spaced by 18.4 GHz (= $2B$) were clearly confirmed within a frequency range of 1 THz in Figs. 6(b) and (c) in the same manner as Fig. 6(a). We could assign 44 groups to $J = 10$ around 0.20 THz to $J = 53$ around 0.98 THz in Fig. 6(c), indicating the capacity to probe multiple absorption lines simultaneously. We confirmed that the noise level at higher frequency in Fig. 6(b) was larger than that in Fig. 6(c). Furthermore, the spectral envelope of the peak absorbance in Fig. 6(b) was less uneven than that in Fig. 6(c). This implies that the adaptive sampling method improves the accuracy of the spectral sampling point and hence enhances the signal-to-noise ratio. From these results, we can conclude that the adaptive sampling, dual THz comb spectroscopy has a high potential for high accuracy, high resolution, and broadband THz spectroscopy even though free-running dual lasers are used.

## 4. Discussions

In the proposed method, the non-linearity of the frequency axis in the mode-resolved THz comb spectrum was effectively corrected by the data acquisition in synchronization with the adaptive sampling clock. On the other hand, the absolute frequency accuracy in the spectrum is not secured because $f_{rep1}$ fluctuates every moment due to free-running operation. The fluctuation of $f_{rep1}$ in the free-running laser used here was 50 mHz at a gate time of 200 ms, corresponding to a single scan time for one temporal waveform (= $1/f_{offset}$). Since this frequency fluctuation is multiplied by $m$, the mode-resolved THz comb spectrum includes the fluctuation of 500 Hz at 1 THz, corresponding to the frequency accuracy of $5 \times 10^{-10}$. Since this fluctuation is much smaller than frequency spacing and linewidth of the comb modes, it will be negligible in dual THz comb spectroscopy.

On the other hand, the spectral resolution in the dual comb spectroscopy is determined by the linewidth of comb mode because the linearity of the frequency scale is recovered by the adaptive sampling method. The come-mode linewidth of 10 MHz in the inset of Fig. 4(e) was determined by the time window size of 100 ns in Fig. 4(c). On the other hand, the minimal limit of the comb-mode linewidth is determined by the relative linewidth of mode between dual THz combs, which is expected to 100

Hz from the result of Fig. 2(c). Therefore, the present system has further linewidth gains that can be implemented if the time window size is further expanded.

5. Conclusions

To promote the wide use of dual THz comb spectroscopy in THz spectroscopic applications, we developed the adaptive sampling method for the dual THz comb spectroscopy with free-running dual femtosecond lasers. A narrow-linewidth, beat signal between dual THz comb modes at 4 THz was extracted by dual THz-comb-referenced THz spectrum analysers and electrical signal processing, and then was used as the adaptive sampling clock in the data acquisition board. As a result, the non-linearity in time scale, induced by the laser timing jitter, was well corrected, and the distinct comb mode was recovered in mode-resolved THz comb spectrum. The demonstrated results did not show only implementation of dual THz comb spectroscopy with free-running dual lasers but also implied the superiority of its spectroscopic performance over the previous dual THz comb spectroscopy with stabilized dual lasers. Although the discrete spectral distribution of THz comb mode limits the spectral sampling interval to $f_{rep1}$ rather than the comb mode linewidth, the combination of the adaptive sampling dual THz comb spectroscopy with the spectral interleaving technique [14] will reduce the spectral sampling interval to the come-mode linewidth. In this case, the benefit of the distinct comb mode [see the inset of Fig. 4 (e)] will be highlighted. Furthermore, the application of the adaptive sampling method should be possible to implement it in ASOPS-THz-TDS [22-25] with free-running lasers, which is more general method for THz spectroscopy than the dual THz comb spectroscopy. The proposed method will lower the threshold of the practical use for dual THz comb spectroscopy, and hence accelerate its applications in real world.

A. Methods

A1. Experimental setup

We used dual mode-locked Er-doped fibre lasers (ASOPS TWIN 100 with P100, Menlo Systems; centre wavelength $\lambda_c$ = 1550 nm, pulse duration $\Delta\tau$ = 50 fs, mean power $P_{mean}$ = 250 mW) for the adaptive sampling dual THz comb spectroscopy (see Fig. 3). The individual repetition frequencies of them ($f_{rep1}$ = 100,000,000 Hz, and $f_{rep2}$ = $f_{rep1}$+$f_{offset}$ = 100,000,050 Hz) and the frequency offset between them ($f_{offset}$ = $f_{rep2}$–$f_{rep1}$ = 50 Hz) were not stabilized. After wavelength conversion of the two laser beams by second-harmonic generation with periodically poled lithium niobate (PPLN) crystals, pulsed THz radiation was emitted from a dipole-shaped, low-temperature-grown (LTG), GaAs photoconductive antenna (PCA1) triggered by pump light ($\lambda_c$ = 775 nm, $\Delta\tau$ = 80 fs, $P_{mean}$ = 19 mW), passed thorough a low-pressure gas cell (length = 500 mm, diameter = 40 mm), and was then detected by another dipole-shaped LTG GaAs photoconductive antenna (PCA2) triggered by probe light ($\lambda_c$ = 775 nm, $\Delta\tau$ = 80 fs, $P_{mean}$ = 9 mW). The optical path, in which the THz beam propagated except for the part in the gas cell, was purged with dry nitrogen gas to avoid absorption by atmospheric moisture. One of two additional output lights from the two lasers was fed into an adaptive-sampling-clock generator to generate the adaptive sampling clock in THz region [see Fig. 2]. The other of them was delivered to a sum-frequency-generation cross-correlator (SFG-XC). The resulting SFG signal was used to generate a time origin signal in the ASOPS measurement. After amplification with a current preamplifier (AMP, bandwidth = 1 MHz, gain = $4\times10^6$ V/A), the temporal waveform of the output current from PCA2 was acquired with a digitizer (sampling rate = $2\times10^6$ samples/s, resolution = 20 bit) by using the SFG-XC's output as a trigger signal and the adaptive-sampling-clock generator's output as a clock signal. Then, the time scale of the observed signal was multiplied by TMF (= $f_{rep1}/f_{offset}$ = 2,000,000). This sampling rate and this TMF enabled us to measure the temporal waveform of 10 consecutive THz pulses at a sampling interval of 100 fs with a time window of 100 ns.

**Acknowledgements**

This work was supported by Collaborative Research Based on Industrial Demand from the Japan Science and Technology Agency, Grants-in-Aid for Scientific Research No. 26246031 from the Ministry of Education, Culture, Sports, Science, and Technology of Japan. We also gratefully acknowledge financial support from the Renovation Centre of Instruments for Science Education and Technology at Osaka University, Japan.

**Author Contributions**

T. Y conceived the idea and wrote the paper. R. I., Y.-D. H, K. H., F. H., and Y. S. built the instrument, performed the experiment, and analysed data. K. M. and H. I. contributed to manuscript preparation and supervised the research.

**Competing Financial Interests statement**

The authors declare no competing financial interests.

**Figure Legends**

Fig. 1. Principle of operation. (a) Signal chart of dual comb spectroscopy in THz and RF regions. (b) Comparison of the data acquisition between the constant sampling method and the adaptive sampling method.

Fig. 2. (a) Extraction of beat signal between dual THz comb modes using dual THz-comb-referenced spectrum analysers. (b) Experimental setup of dual THz-comb-referenced spectrum analysers. (c) Extracted beat signal between dual THz comb modes around 4 THz.

Fig. 3. Experimental setup.

Fig. 4. Comparison of the accumulated temporal waveforms among (a) constant sampling method with free-running dual lasers, (b) constant sampling method with stabilized dual lasers, and (c) adaptive sampling method with free-running dual lasers. Comparison of the mode-resolved THz comb spectrum between (d) constant sampling method with stabilized dual lasers obtained by FT of Fig. 4(b) and (e)

adaptive sampling method with free-running dual lasers obtained by FT of Fig. 4(c).

Fig. 5. Comparison of absorbance spectrum of the water rotational-transition ($1_{10} \leftarrow 1_{01}$) between (a) constant sampling method with stabilized dual lasers and (b) adaptive sampling method with free-running dual lasers.

Fig. 6. Comparison of absorbance spectrum of low-pressure $CH_3CN$ gas among (a) spectral database, (b) constant sampling method with stabilized dual lasers, and (c) adaptive sampling method with free-running dual lasers.

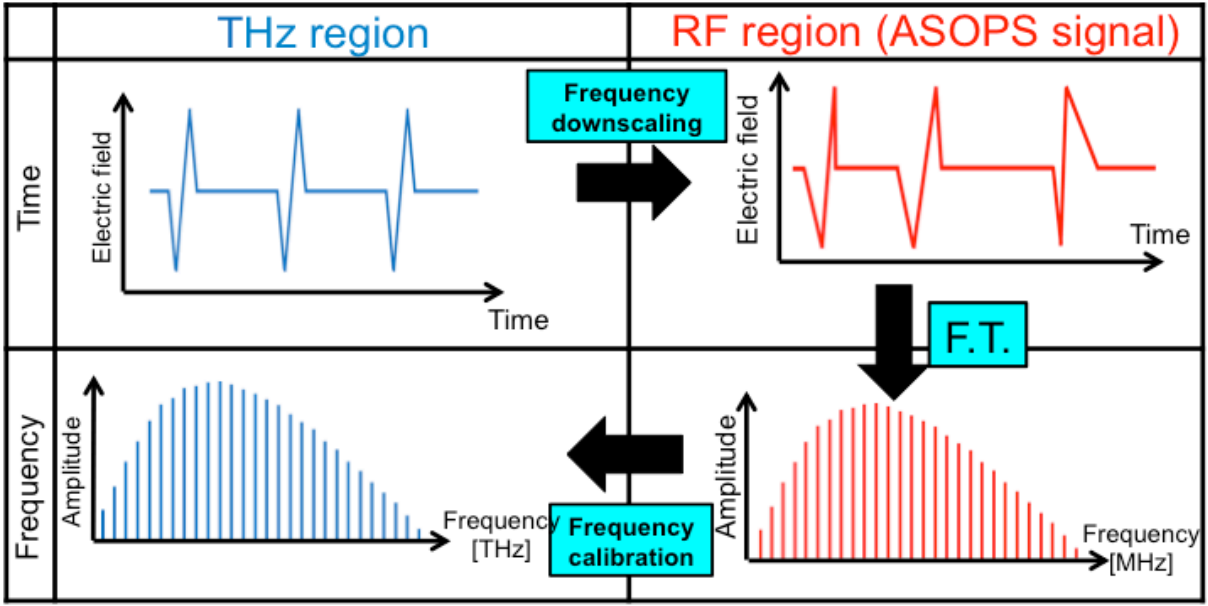

(a)

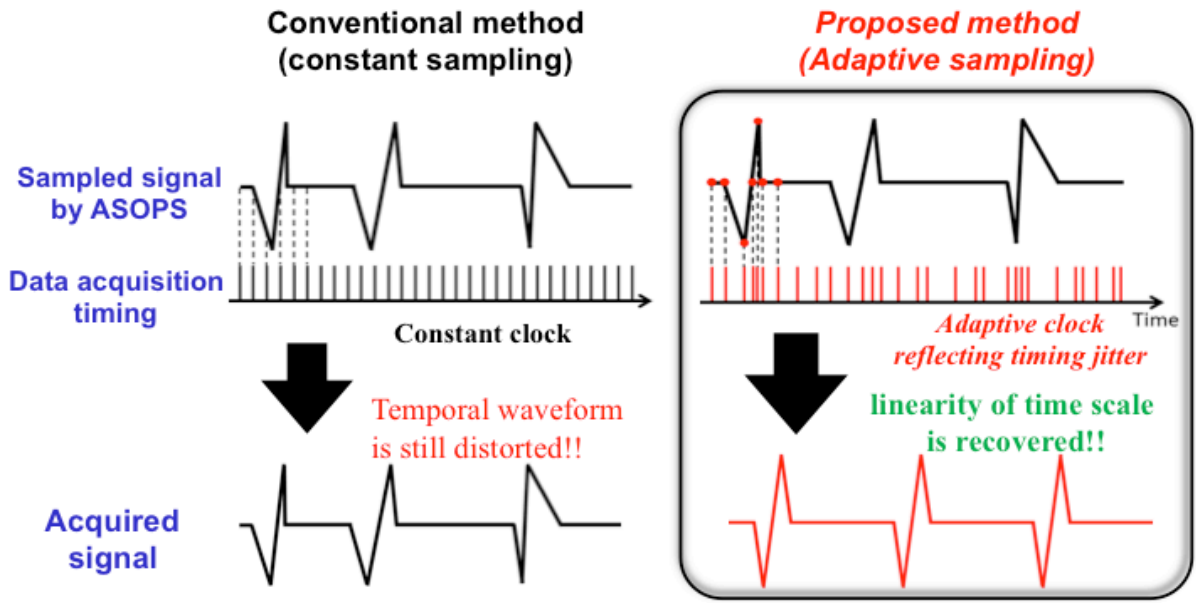

(b)

Fig. 1.

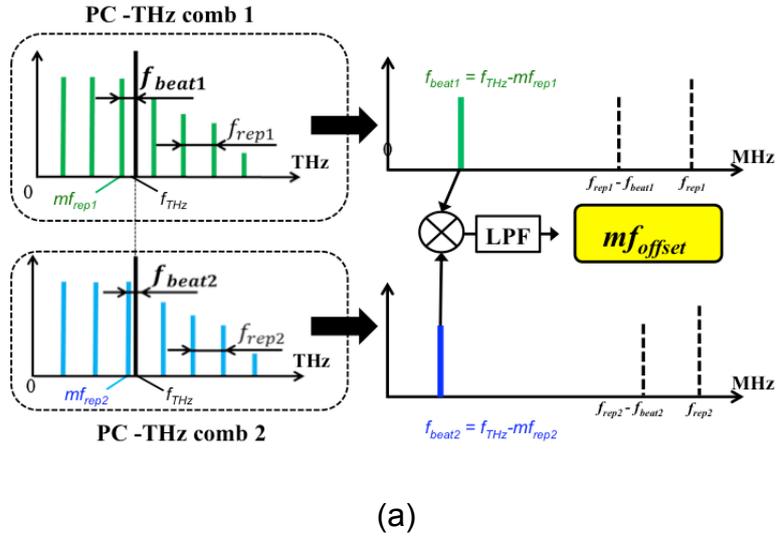

(a)

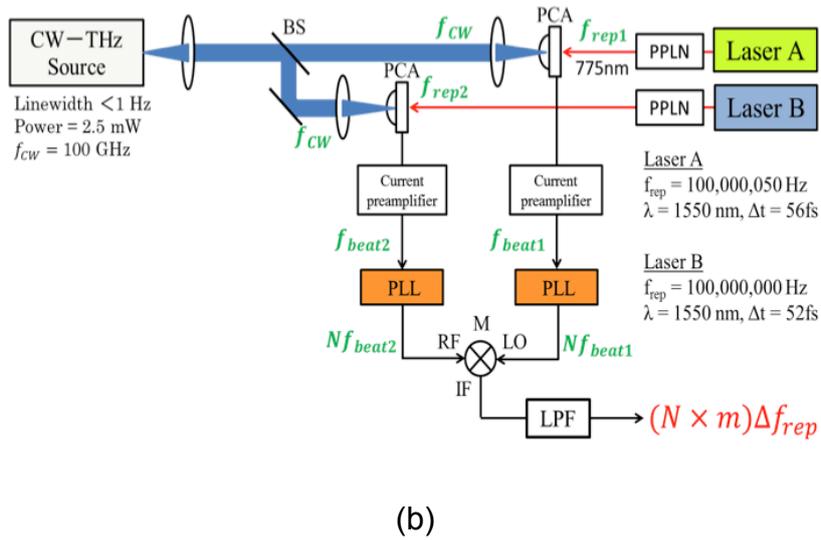

(b)

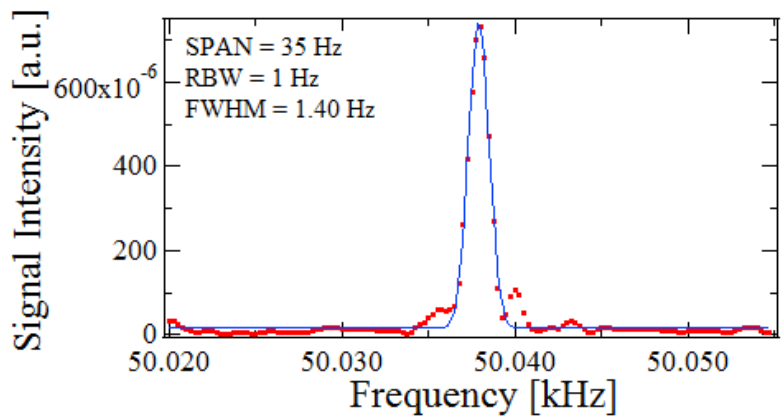

(c)

Fig. 2.

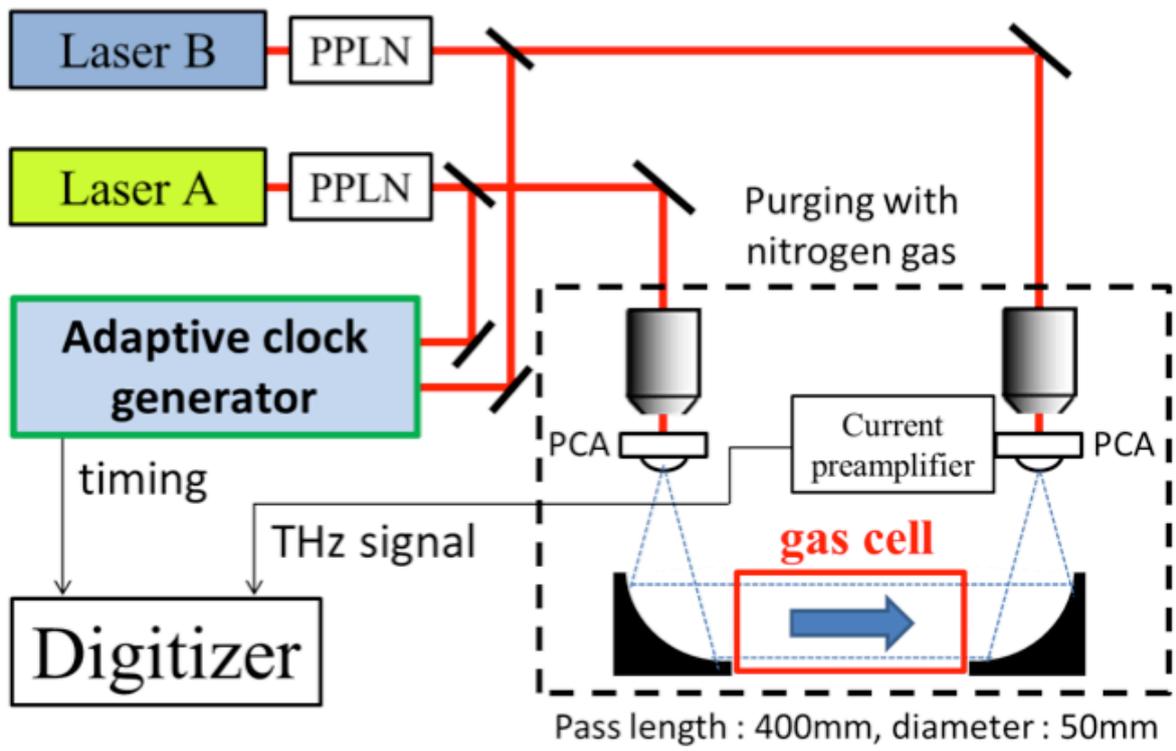

Fig. 3.

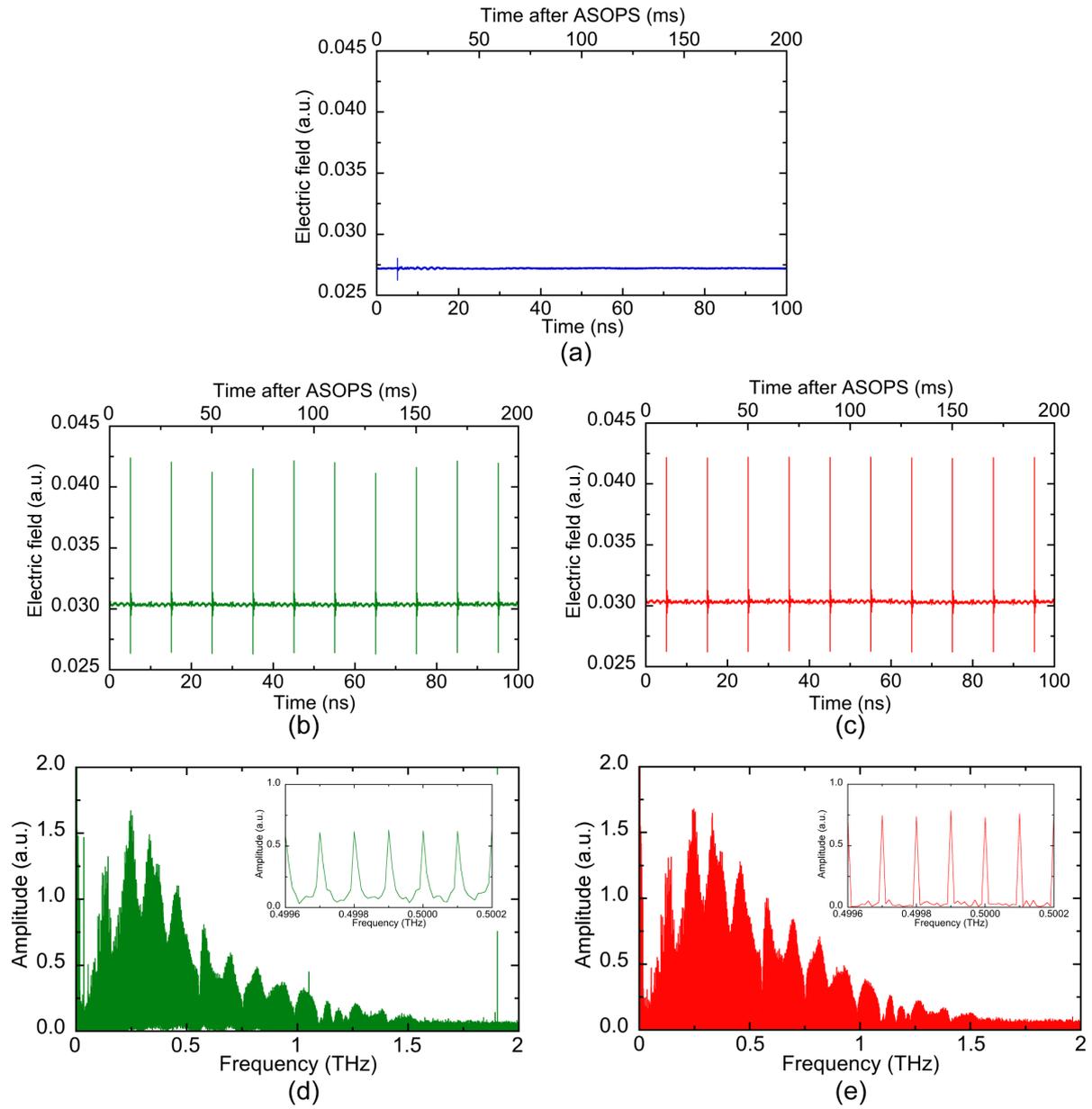

Fig. 4.

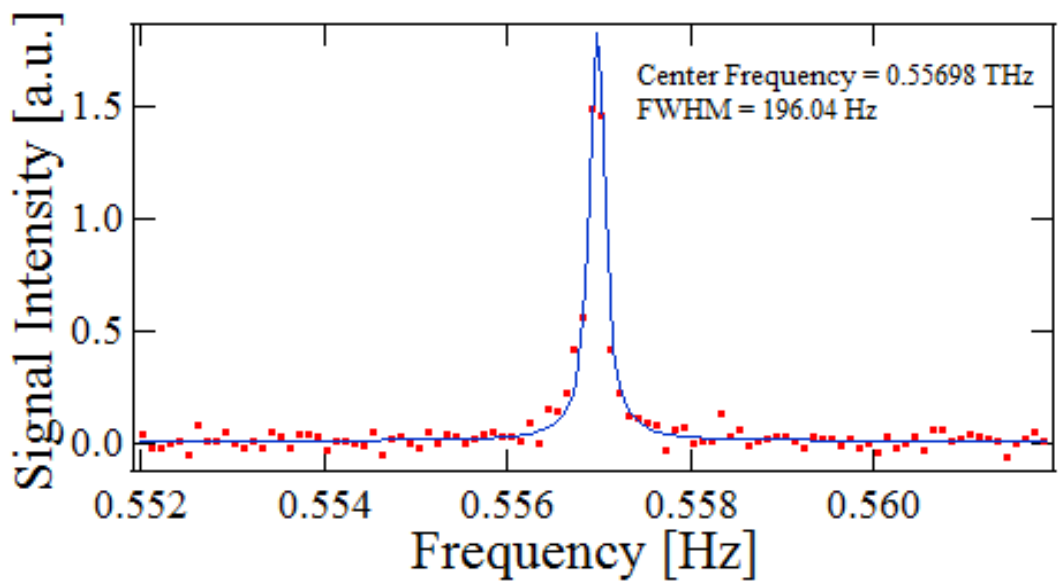

(a)

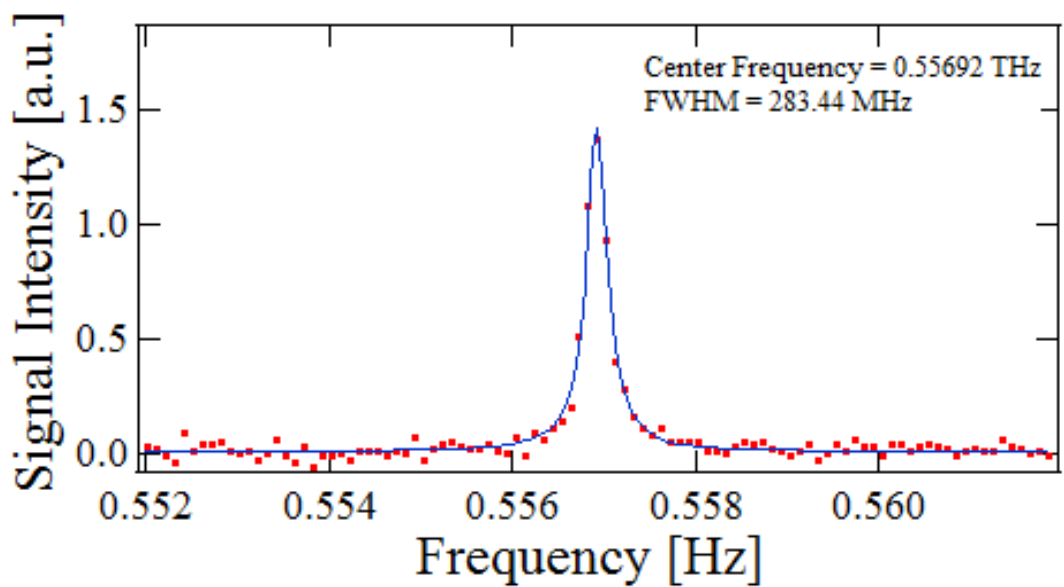

(b)

Fig. 5.

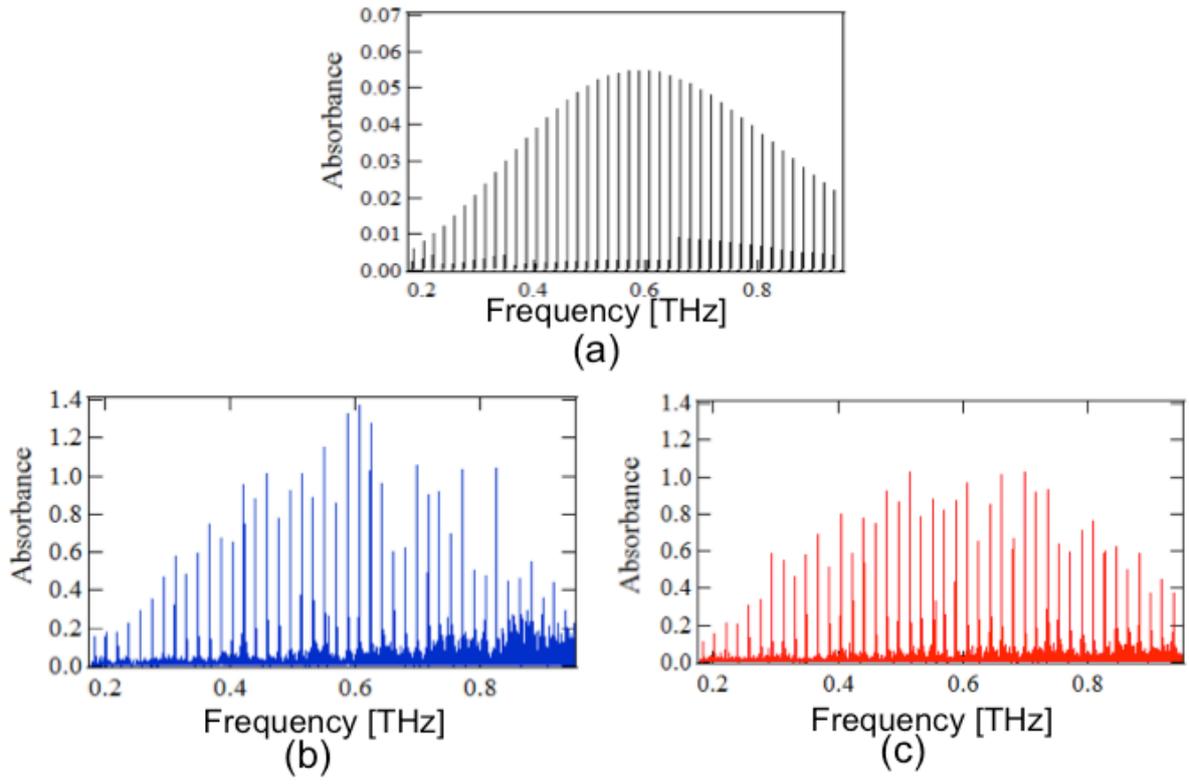

Fig. 6.